\documentclass[11pt]{article}

\usepackage[preprint]{acl}

\usepackage{times}
\usepackage{latexsym}

\usepackage[T1]{fontenc}

\usepackage[utf8]{inputenc}

\usepackage{microtype}

\usepackage{inconsolata}

\usepackage{listings}
\usepackage{graphicx}
\usepackage[ruled,vlined]{algorithm2e} 

\usepackage{booktabs}
\usepackage{multirow}
\usepackage{amsmath}
\usepackage{makecell}
\usepackage[most]{tcolorbox}
\usepackage{listings}
\usepackage[table]{xcolor}
\tcbset{
  promptstyle/.style={
    enhanced,
    breakable,
    colback=gray!10,        
    colframe=black!40,      
    colbacktitle=gray!20,   
    coltitle=black,         
    boxrule=0.4pt,          
    arc=2pt,                
    outer arc=2pt,
    left=5pt, right=5pt, top=5pt, bottom=5pt,
    fonttitle=\bfseries,    
    title={#1}              
  }
}
\usepackage{fvextra}

\title{PyraVid: Hierarchical Multimodal Memory for Long-Horizon Video Reasoning}

\author{
 \textbf{Sikuan Yan\textsuperscript{*1,2,3}},
 \textbf{Sicheng Dong\textsuperscript{*4}},
 \textbf{Haotong Wang\textsuperscript{4}},
 \textbf{Ercong Nie\textsuperscript{1}},
\\
 \textbf{Yilun Liu\textsuperscript{1}},
 \textbf{Jinhe Bi\textsuperscript{1}},
 \textbf{Yingjie Xu\textsuperscript{4}},
 \textbf{Susanna Schwarzmann\textsuperscript{3}},
\\
 \textbf{Riccardo Trivisonno\textsuperscript{3}},
 \textbf{Volker Tresp\textsuperscript{1,2}},
 \textbf{Yunpu Ma\textsuperscript{\dag}\textsuperscript{1,2}}
\\
 \textsuperscript{1}Ludwig Maximilian University of Munich,
 \textsuperscript{2}Munich Center for Machine Learning,
\\
 \textsuperscript{3}Huawei Heisenberg Research Center (Munich),
 \textsuperscript{4}Technical University of Munich
\\
 \small{
    \href{mailto:email@domain}{s.yan@campus.lmu.de}, \href{mailto:email@domain}{cognitive.yunpu@gmail.com}
 }
}

\begin{document}
\maketitle
\renewcommand{\thefootnote}{\fnsymbol{footnote}}
\footnotetext[1]{Equal contribution.}
\footnotetext[2]{Corresponding author.}
\renewcommand{\thefootnote}{\arabic{footnote}}
\begin{abstract}
Memory has become an increasingly important component of agentic systems, as these systems are expected to reason over long-term experience. However, prior work has largely focused on unimodal memory, leaving multimodal memory relatively underexplored despite its central role in real-world applications. Compared with unimodal settings, multimodal memory introduces additional challenges, including heterogeneous input integration, person-centric information alignment, and evidence aggregation across different granularities. We present PyraVid, a hierarchical multimodal memory framework inspired by Event Segmentation Theory from cognitive science. PyraVid organizes long videos into a coarse-to-fine pyramid structure, enabling structured memory access and effective evidence aggregation. It further supports structure-guided memory expansion with pruning, allowing the retrieval of related events with strong causal connectivity but low semantic similarity while reducing noise. Experiments on multiple long-video understanding benchmarks show that PyraVid consistently improves performance across datasets, model scales, and question types, highlighting the effectiveness of hierarchical multimodal memory for long-horizon reasoning.
\end{abstract}

\begin{figure*}[t]
\centering
\includegraphics[width=1\textwidth]{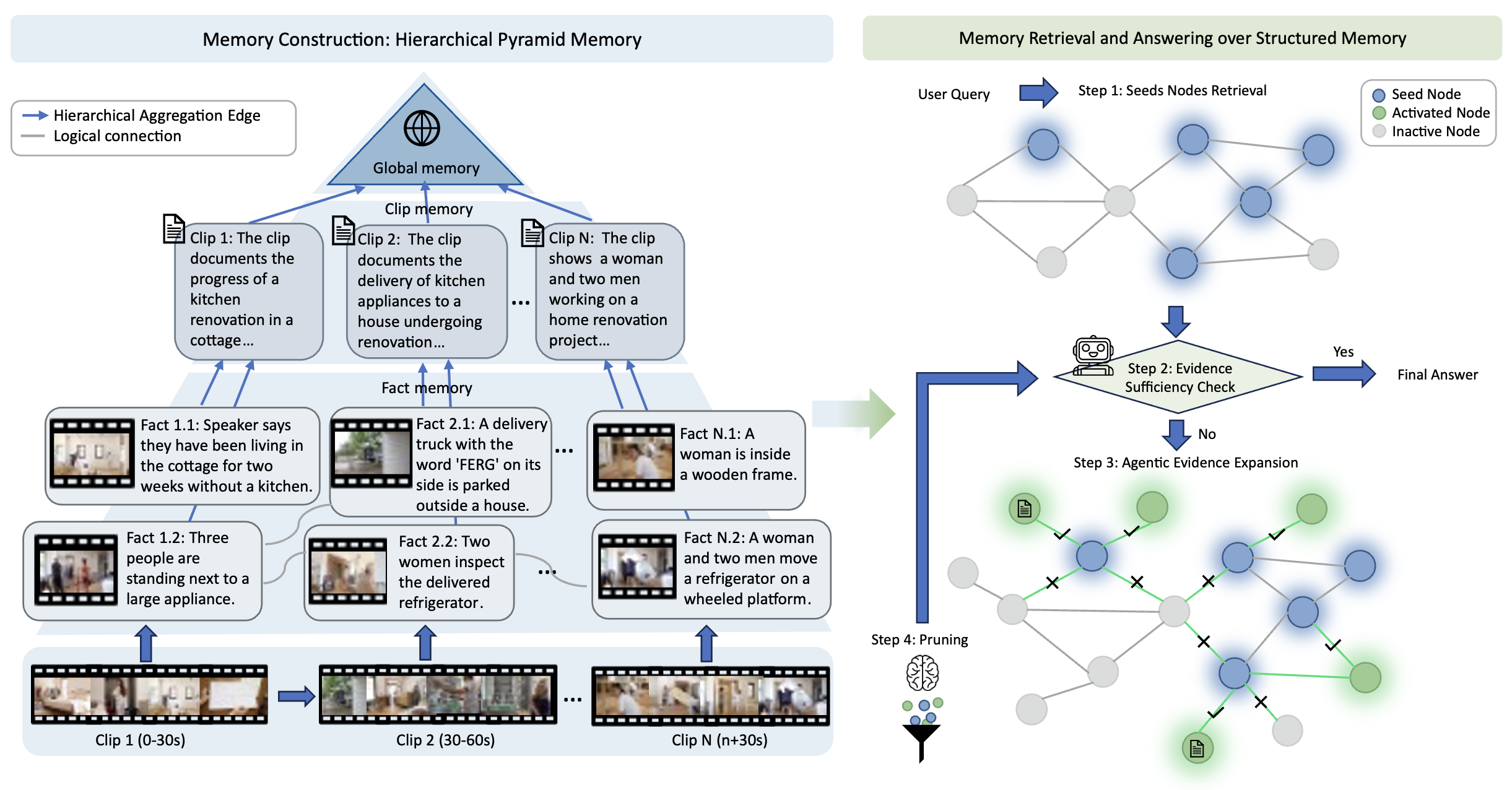} 
\caption{
Overview of PyraVid. Left: PyraVid organizes streaming video into a hierarchical pyramid memory with fact, clip, and global memory connected by structured links. Right: given a query, PyraVid retrieves seed nodes, expands to related evidence over the memory graph, prunes irrelevant nodes, and produces the final answer.
}
\label{fig:main}
\end{figure*}

\section{Introduction}

Agent memory has become an increasingly important topic in recent research, motivated by the need for agents to reason over long-term experience. A common paradigm is to maintain an external memory bank that is continuously updated as new information arrives, and to retrieve relevant entries at inference time to support decision-making or response generation~\cite{chhikara2025mem0buildingproductionreadyai, li2025memosoperatingmemoryaugmentedgeneration}. Existing work has made substantial progress on text-based memory systems, including memory life-cycle management~\cite{zhong2023memorybankenhancinglargelanguage, chhikara2025mem0buildingproductionreadyai, li2025memosoperatingmemoryaugmentedgeneration}, graph-based memory organization~\cite{wang2025mirixmultiagentmemoryllmbased, rasmussen2025zeptemporalknowledgegraph}, and hierarchical memory structures~\cite{li2025camconstructivistviewagentic, hu2024hiagenthierarchicalworkingmemory}. 

Real-world environments are inherently multimodal, involving visual, auditory, and temporal signals. This can be naturally formulated as an online video understanding setting, in which a multimodal agent must process an incoming video stream incrementally, retain what has happened over time, and organize information from different modalities for future reasoning. Compared with text-only memory, this setting introduces additional challenges, because relevant evidence may be expressed through heterogeneous modalities and distributed across distant events. Existing multimodal memory systems~\cite{mao2025multiragmultimodalretrievalaugmentedgeneration, lin2025hippommhippocampalinspiredmultimodalmemory, m3agent} typically represent memory as a flat or weakly structured collection of entries. As a result, they provide limited support for coordinating information across different levels of abstraction, even though long-horizon video understanding often requires combining fine-grained observations with higher-level event context. In addition, not all relevant evidence can be directly retrieved through semantic similarity to the query. For example, given the query "Why did the person return to the kitchen?", the most informative evidence may not be the return action itself, but an earlier event "leaving the kettle on the stove". Although this event shares little semantic overlap with the query, it is causally essential for answering it. These limitations highlight the need for a structured multimodal memory system that can organize video experience across levels of abstraction and support evidence composition during reasoning.

Inspired by Event Segmentation Theory~\cite{zacks2007event} in cognitive science, which suggests that humans parse continuous experience into meaningful events, often across multiple temporal scales, we propose PyraVid, a hierarchical multimodal memory framework for online video processing. PyraVid organizes streaming video into a coarse-to-fine memory pyramid with explicit temporal and causal links, and performs inference by iteratively expanding and pruning evidence over this structure. We evaluate PyraVid on four long-video understanding benchmarks against six representative baselines. Results show that PyraVid consistently outperforms prior approaches across benchmarks and evaluation settings. Further ablations validate the effectiveness of both the overall framework and its key components.

Our contributions are threefold:
(1) We introduce PyraVid, a hierarchical multimodal memory framework for online long-video understanding, which organizes streaming video into a coarse-to-fine memory pyramid spanning fact-level observations, clip-level event abstractions, and global-level understanding.
(2) We propose a structure-guided reasoning mechanism that explicitly exploits the memory hierarchy by expanding relevant evidence through memory links and pruning irrelevant nodes, enabling effective composition of evidence across temporal spans and levels of abstraction.
(3) We conduct extensive experiments on four benchmarks, showing that PyraVid consistently outperforms strong baselines across benchmarks, model scales, and question types, with ablations further validating the effectiveness of the proposed design.

\section{Related Work}
\subsection{Agent Memory}

Agent memory has become a key component of modern agent systems, addressing the limited context window of large language models by enabling persistent access to long-term knowledge. Early work primarily focused on text-based memory systems~\cite{chhikara2025mem0buildingproductionreadyai, li2025memosoperatingmemoryaugmentedgeneration, xu2025amemagenticmemoryllm}, establishing core mechanisms for memory construction, maintenance, and retrieval. More recent studies have further improved memory management and retrieval capabilities~\cite{yan2026memoryr1enhancinglargelanguage, wang2025memalphalearningmemoryconstruction, yue2026memtdensifyingrewardslonghorizon}, while also introducing more structured memory representations. For example, A-MEM~\cite{xu2025amemagenticmemoryllm} adopts an agentic memory design inspired by Zettelkasten, whereas CAM~\cite{li2025camconstructivistviewagentic} and HiAgent~\cite{hu2024hiagenthierarchicalworkingmemory} propose a hierarchical working memory framework that organizes past trajectories around subgoals, balancing compact summarization with retrieval efficiency.

Despite this progress, most existing agent memory systems remain limited to textual representations. Such designs are insufficient for long-horizon video understanding, which involves heterogeneous inputs and requires reasoning over temporally extended events, cross-modal alignment, and visually grounded evidence. M3-Agent~\cite{m3agent} represents an important early step toward multimodal memory in this setting. However, it does not fully exploit memory structure or visual information during inference. This limitation motivates the development of a multimodal memory framework that supports structure-aware retrieval and reasoning.

\subsection{Online Video Understanding}

Online video understanding studies how systems continuously interpret streaming video while retaining useful information over time. Recent advances in video large language models have improved long-video reasoning by extending context windows or scaling multimodal encoders~\cite{bai2025qwen3, comanici2025gemini}. However, these approaches typically process long videos as enlarged offline inputs, leading to high computational cost and limited flexibility in streaming settings.

To improve efficiency, prior work has explored compressed video representations, such as visual token reduction and sparse frame selection~\cite{li2024videochat,wang2025videotree}. Although these methods reduce processing cost, they often discard fine-grained details or weaken temporal continuity. M3-Agent~\cite{m3agent} takes a step toward online long-video understanding by introducing a multimodal long-term memory that stores episodic and semantic information in an entity-centric graph. However, during inference, its memory is primarily consumed in textual form, which leaves visual evidence underutilized and limits the benefits of the underlying memory structure. These limitations suggest that long-horizon video understanding requires memory mechanisms that can organize and retrieve evidence across multiple levels of granularity. Our work addresses this challenge with a hierarchical multimodal memory framework that preserves fact-, clip-, and global memories, and their structural relations, enabling structure-aware retrieval and reasoning over long videos.

\section{PyraVid}
Long-horizon multimodal video understanding is challenging because relevant evidence may be distributed across temporally distant segments, heterogeneous modalities, and multiple levels of granularities. Answering a query often requires coherently aggregating fine-grained observations, high-level semantic information, and long-range contextual cues. To address this challenge, we present PyraVid, a hierarchical multimodal memory framework that is designed to process streaming video inputs online, organize extracted information into a coarse-to-fine memory pyramid, and exploit this structure during inference for effective evidence aggregation. In the following, Section~\ref{sec:problem_formulation} formalizes the task, Section~\ref{sec:pyravid_memory} introduces the hierarchical memory construction of PyraVid, and Section~\ref{sec:pyravid_answerpart} describes the structure-guided reasoning and answer generation process.

\subsection{Problem Formulation}
\label{sec:problem_formulation}

We consider the problem of online long-video understanding, in which a multimodal agent receives a streaming video input and answers queries based on accumulated observations. Unlike in offline settings, the full video is not accessible at inference time, so the agent must process incoming content on the fly, extract salient information, and maintain a compact memory for future reasoning.

Formally, given a streaming video $\mathcal{V}$, the agent constructs a memory $\mathcal{M}$ while observing the stream, which may contain multimodal signals such as visual content, speech, and temporally localized events. At test time, given a query $q$, the agent retrieves relevant evidence from $\mathcal{M}$ to produce an answer $a$ without revisiting the original video. The central challenge, therefore, is to design a memory representation that preserves informative observations, supports long-range temporal reasoning, and enables efficient retrieval over extensive evidence.

\subsection{Pyramid Memory Structure}
\label{sec:pyravid_memory}

We organize long-video memory as a hierarchical pyramid spanning multiple levels of abstraction. The key intuition is that long-horizon reasoning requires access to information at different granularities, ranging from grounded local observations to high-level global understanding. Accordingly, PyraVid maintains three complementary levels of memory: \emph{fact memory}, \emph{clip memory}, and \emph{global memory}.

\paragraph{Hierarchical Memory Definition}
Formally, the memory state at time step $t$ is defined as
\begin{equation}
\mathcal{M}^{(t)} = \langle \mathcal{M}_{\text{global}}^{(t)}, \mathcal{M}_{\text{clip}}^{(t)}, \mathcal{M}_{\text{fact}}^{(t)} \rangle,
\end{equation}
where $\mathcal{M}_{\text{fact}}^{(t)}$ stores fine-grained episodic observations describing what occurs in the video, $\mathcal{M}_{\text{clip}}^{(t)}$ summarizes events within local temporal segments and captures higher-level semantic information, and $\mathcal{M}_{\text{global}}^{(t)}$ maintains an evolving global understanding of the video to provide contextual guidance for reasoning. The hierarchical memory bank is updated incrementally as the video stream is processed online.

\paragraph{Fact Memory}
Fact memory stores fine-grained multimodal observations extracted from the incoming video stream, capturing episodic evidence about events and states over time. As the video is processed online, PyraVid incrementally augments the fact memory with newly extracted fact nodes from the current clip:
\begin{equation}
\mathcal{M}_{\text{fact}}^{(t)} = \mathcal{M}_{\text{fact}}^{(t-1)} \cup \{ m_i^{(t)} \}_{i=1}^{N_t},
\end{equation}
where $\{ m_i^{(t)} \}_{i=1}^{N_t}$ denotes the set of fact memories extracted at time step $t$. Each fact memory is represented as
\begin{equation}
m_i = \langle \tau_i, v_i, x_i, \mathcal{L}(m_i) \rangle,
\end{equation}
where $\tau_i$ denotes temporal information, $v_i$ denotes the associated visual evidence, and $x_i$ is a grounded textual description that encodes fine-grained observations from the segment, such as person-related information, scene context, and event-level activities. $\mathcal{L}(m_i)$ denotes the set of links from the current node to other structurally or logically related nodes, as detailed in Section~\ref{sec:link_construction}. This representation separates factual content from structural connectivity, allowing fact memories to remain expressive while supporting compositional aggregation and structure-aware reasoning.

\paragraph{Clip Memory}
Clip memory provides a compact semantic representation of a local temporal segment. For each processed clip, PyraVid generates a clip-level memory that summarizes the key events and contextual information within the segment. Unlike fact memory, which captures fine-grained episodic observations, clip memory abstracts them into a higher-level semantic description of the clip. These clip memories serve as an intermediate representation between fact memory and global memory, enabling efficient reasoning over medium-range temporal spans.

\paragraph{Global Memory}
Global memory maintains an evolving high-level representation of the entire video. After processing each clip, the global memory is updated as
\begin{equation}
\mathcal{M}_{\text{global}}^{(t)} = \mathcal{U}_{\text{global}}(\mathcal{M}_{\text{global}}^{(t-1)}, \mathcal{M}_{\text{clip}}^{(t)}),
\end{equation}
where $\mathcal{U}_{\text{global}}(\cdot)$ is an incremental update function that integrates newly formed clip-level memories into the global representation.

Global memory captures long-range context and high-level video semantics, providing contextual guidance for retrieval and reasoning in queries that require a holistic understanding of the video.

\paragraph{Link Construction}
\label{sec:link_construction}

The three memory levels in PyraVid are connected through structured links rather than maintained as isolated representations. These links operate both across granularity levels, enabling hierarchical aggregation, and within the same level, capturing logical dependencies. PyraVid first constructs hierarchical links across the memory pyramid: each fact memory is connected to its corresponding clip memory, and each clip memory is linked to the global memory, forming a bottom-up pathway from grounded observations to global understanding. In addition, PyraVid constructs relational links among fact memories. Formally, each fact memory $m_i$ is associated with a set of outgoing links $\mathcal{L}(m_i)$ generated using a sparse linking strategy. For each $m_i$, the system retrieves a small set of semantically similar candidate facts and establishes links only when clear logical relations are identified. This design keeps the fact graph compact and reduces retrieval noise. Together, these hierarchical and relational links form a structured multimodal memory graph that supports structure-aware evidence expansion during inference, as described in Section~\ref{sec:pyravid_answerpart}.

\subsection{Structure-Guided Reasoning over Hierarchical Memory}
\label{sec:pyravid_answerpart}

Given a query $q$, PyraVid performs inference over the hierarchical memory by iteratively retrieving relevant nodes, assessing their sufficiency, expanding to related evidence, and pruning irrelevant information. Rather than relying on a single retrieval step, the system progressively gathers supporting evidence from the structured memory graph until sufficient context is obtained for answer generation.

\paragraph{Seed Retrieval}
PyraVid first retrieves a small set of seed memory nodes that are semantically relevant to the query using embedding-based retrieval. Given the fact memory set $\mathcal{M}_{\text{fact}}$, the initial seed set is defined as
\begin{equation}
\mathcal{C}^{(0)} = \text{TopK}_{m_i \in \mathcal{M}_{\text{fact}}} \; \text{sim}(q, m_i),
\end{equation}
where $\text{sim}(\cdot)$ measures the similarity between the query and the textual content of each fact memory.

\paragraph{Evidence Sufficiency Assessment}
Starting from the seed set $\mathcal{C}^{(0)}$, PyraVid invokes a model to assess whether the currently retrieved evidence is sufficient to answer the query. At iteration $r$, let $\mathcal{C}^{(r)}$ denote the current evidence context. Given the query and the current evidence, the model outputs a candidate answer together with a sufficiency indicator:
\begin{equation}
a^{(r)} = f_{\text{reason}}(q, \mathcal{C}^{(r)}),
\end{equation}
where $a^{(r)}$ denotes the model output at iteration $r$. If the current evidence is sufficient to answer the query, $a^{(r)}$ is the final answer, and iteration terminates. Otherwise, $a^{(r)}$ is a special signal indicating that additional evidence is required, and PyraVid continues to retrieve and expand the evidence context.

\paragraph{Structure-Guided Expansion}
When the current evidence is insufficient, PyraVid expands the candidate set by traversing the structured links associated with the currently activated memory nodes. Let $\mathcal{A}^{(r)}$ denote the set of active memory nodes at iteration $r$, and let $\mathcal{L}(m_i)$ denote the outgoing structured links of node $m_i$. The expansion step collects neighboring memory nodes as
\begin{equation}
\mathcal{E}^{(r)} = \bigcup_{m_i \in \mathcal{A}^{(r)}} \mathcal{L}(m_i),
\end{equation}
where the traversed links may include hierarchical and causal connections. 

\paragraph{Agent-Based Pruning}
The expanded candidate set may contain irrelevant or weakly related nodes to the query. To reduce noise, PyraVid employs a pruning agent that evaluates each expanded node conditioned on the query and the current evidence context. For each candidate node $m_i \in \mathcal{E}^{(r)}$, the pruning agent predicts a binary selection decision:
\begin{equation}
z_i^{(r)} = f_{\text{prune}}(q, m_i), \quad z_i^{(r)} \in \{0,1\}.
\end{equation}
The retained nodes form the pruned expansion set:
\begin{equation}
\tilde{\mathcal{E}}^{(r)} = \{ m_i \mid m_i \in \mathcal{E}^{(r)} \;\wedge\; z_i^{(r)} = 1 \}.
\end{equation}
These retained nodes are then incorporated into the current evidence context:
\begin{equation}
\mathcal{C}^{(r+1)} = \mathcal{C}^{(r)} \cup \tilde{\mathcal{E}}^{(r)}.
\end{equation}

\paragraph{Iterative Reasoning and Answer Generation}
The assessment, expansion, and pruning steps are repeated until the model determines that the evidence is sufficient or a maximum number of iterations $R$ is reached. This iterative inference strategy allows PyraVid to progressively aggregate distributed evidence from the structured memory graph while avoiding the excessive noise introduced by unrestricted expansion. As a result, PyraVid enables efficient long-horizon reasoning over multimodal video memory.

\begin{table*}[t]
\centering
\footnotesize
\setlength{\tabcolsep}{2pt}
\renewcommand{\arraystretch}{1.2}

\resizebox{\textwidth}{!}{
\begin{tabular}{lcccccc cccccc cc}
\toprule
\multirow{2}{*}{\textbf{Model}} &
\multicolumn{6}{c}{\textbf{M3-Bench-robot}} &
\multicolumn{6}{c}{\textbf{M3-Bench-web}} &
\multirow{2}{*}{\textbf{VM(L)}} &
\multirow{2}{*}{\textbf{LVB}}  \\
\cmidrule(lr){2-7}\cmidrule(lr){8-13}
& \textbf{MDR} & \textbf{MHR} & \textbf{CMR} & \textbf{PU} & \textbf{GKE} & \textbf{ALL}
& \textbf{MDR} & \textbf{MHR} & \textbf{CMR} & \textbf{PU} & \textbf{GKE} & \textbf{ALL}
& & \\
\midrule

\multicolumn{15}{c}{\textbf{Socratic Model}} \\
Qwen3-VL-8B-Instruct     & 25.9 & 25.0 & 24.2 & 28.5 & 25.5 & 25.0 & 46.5 & 26.5 & 27.3 & 46.7 & 29.4 & 35.7 & 41.46 & 45.03 \\
Gemini-2.0-Flash  & 30.4 & 35.0 & 30.8 & 34.4 & 28.7 & 31.5 & 52.5 & 30.6 & 34.1 & 46.0 & 51.0 & 44.9 & 62.60 & 49.25 \\
\addlinespace[2pt]
\midrule

\multicolumn{15}{c}{\textbf{Online Video Understanding Methods}} \\
\addlinespace[1pt]
MovieChat      & 13.3 & 9.8 & 12.2 & 15.7 & 7.0 & 11.2 & 12.2 & 6.6 & 12.5 & 17.4 & 11.1 & 12.6 & 19.4 & 22.5  \\
MA-LMM         & 25.6 & 23.4 & 22.7 & 39.1 & 14.4 & 24.4 & 26.8 & 10.5 & 22.4 & 39.3 & 15.8 & 24.3  & 17.3  & 30.0 \\
Flash-VStream  & 21.6 & 19.4 & 19.3 & 24.3 & 14.1 & 19.4 & 24.5 & 10.3 & 24.6 & 32.5 & 20.2 & 23.6 & 25.0 & 42.0 \\
\addlinespace[2pt]
\midrule
         
\multicolumn{15}{c}{\textbf{Agent Method}} \\
\addlinespace[1pt]
M3-Agent(32B RL)       & 32.8 & 29.4 & 31.2 & 43.3 & 19.1 & 30.7 & 45.9 & 28.4 & 44.3 & 59.3 & \textbf{53.9} & 48.9 & 55.3 & 49.3  \\
\rowcolor{gray!15}
PyraVid(8B)        & 43.0 & 35.0 & 39.1 & 56.1 & 26.8 & 40.9 & 52.5 & 32.7 & 43.2 & 67.9 & 45.1 & 51.1 & 59.5 & 50.6 \\
\rowcolor{gray!15}
PyraVid(32B)        & \textbf{47.8} & \textbf{50.0} & \textbf{44.4} & \textbf{60.9} & \textbf{36.3} & \textbf{46.7} & \textbf{55.4} & \textbf{42.9} & \textbf{54.5} & \textbf{70.8} & 51.0 & \textbf{56.3} & \textbf{69.1} & \textbf{58.5}\\
\bottomrule
\end{tabular}
}
\caption{Main results on four long-video understanding benchmarks. For M3-Bench-robot and M3-Bench-web, we report LLM-as-a-judge scores on six dimensions: Multi-Detail Reasoning (MDR), Multi-Hop Reasoning (MHR), Cross-Modal Reasoning (CMR), Person Understanding (PU), General Knowledge Extraction (GKE), and the overall average (ALL). For Video-MME (Long) (VM(L)) and LVBench (LVB), we report multiple-choice accuracy.}
\label{table:main results}
\end{table*}

\section{Experiments}
\subsection{Experimental Setup}

\paragraph{Datasets and Metrics}
We evaluate PyraVid on four long-video understanding benchmarks that require reasoning over extended temporal contexts. M3-Bench-robot and M3-Bench-web~\cite{m3agent} are benchmarks for memory-based multimodal video reasoning. The former contains robot-perspective videos recorded in real-world environments, while the latter consists of diverse web-sourced videos annotated with open-ended question answering tasks. Both benchmarks emphasize long-horizon reasoning over multimodal streams, including cross-modal understanding, person-centric reasoning, and knowledge extraction. 

We further evaluate PyraVid on general long-video understanding benchmarks, including Video-MME (Long)~\cite{videomme} and LVBench~\cite{lvbench}, which contain hour-level or otherwise extended videos designed to assess temporal comprehension and reasoning over long visual contexts. Following prior work~\cite{m3agent}, we adopt LLM-as-a-Judge for evaluating M3-Bench-robot and M3-Bench-web, since both benchmarks involve open-ended questions. The prompt used for LLM-as-a-Judge is provided in Appendix~\ref{app:prompts}. Video-MME (Long) and LVBench are multiple-choice benchmarks, and we report accuracy based on whether the predicted option matches the ground-truth answer.

\paragraph{Baselines}
We compare PyraVid against representative baselines from three categories. 
(1) Socratic models formulate long-video understanding as language-based memory construction followed by retrieval-augmented answering. In this setting, multimodal models such as Qwen3-VL-8B-Instruct~\cite{bai2025qwen3} and Gemini-2.0-Flash~\cite{comanici2025gemini} summarize video content into textual memory for downstream question answering. 
(2) Online video understanding methods include MovieChat~\cite{moviechat}, MA-LMM~\cite{malmm}, and Flash-VStream~\cite{flashvstream}. MovieChat employs a sliding-window mechanism with short- and long-term visual memory for interactive long-video understanding. MA-LMM incrementally processes frames and maintains memory banks to model long-range temporal dependencies. Flash-VStream adopts an asynchronous pipeline with its STAR memory to compress and retrieve visual information from streaming videos. 
(3) Agent-based memory systems are represented by M3-Agent~\cite{m3agent}, which constructs multimodal long-term memory from video and audio streams and performs iterative retrieval and reasoning. For fair comparison, we follow the official implementations or default settings of these baselines whenever available.

\paragraph{Implementation Details}
For PyraVid, we use Gemini-2.0-Flash to construct the hierarchical multimodal memory, including fact extraction, clip-level summarization, and global memory updating. Structured links among memory nodes are generated by Qwen3-4B-Instruct. During inference, we employ separate models for memory selection and answer generation. Specifically, Qwen3-8B and Qwen3-32B serve as the selection models in the iterative retrieval process, while Qwen3-8B-VL-Instruct and Qwen3-32B-Instruct are used as the corresponding answer models to generate the final responses. We denote these two configurations as PyraVid(8B) and PyraVid(32B), respectively. Additional implementation details are provided in Appendix~\ref{app:implementation}.


\subsection{Main Results}

Table~\ref{table:main results} presents the main results on four long-video understanding benchmarks: M3-Bench-robot, M3-Bench-web, Video-MME (Long), and LVBench. Overall, PyraVid consistently outperforms all compared baselines across all benchmarks. In particular, PyraVid(32B) achieves the best overall performance on both M3-Bench-robot and M3-Bench-web, with scores of 46.7 and 56.3, respectively, while also attaining the strongest results on Video-MME and LVBench, scoring 69.1 and 58.5. Even the smaller PyraVid(8B) variant surpasses the strongest prior agent-based baseline, M3-Agent(32B RL), on all reported benchmarks. Compared with Socratic-model baselines and online video understanding methods, PyraVid shows clear advantages in both benchmark-specific subtasks and overall performance, suggesting that hierarchical multimodal memory is more effective for long-horizon reasoning than flat textual memory or feature-based streaming memory. Moreover, the comparison with M3-Agent indicates that the improvement does not stem merely from using a memory module, but from how memory is organized and exploited during inference. These results demonstrate the effectiveness of PyraVid’s pyramid memory structure and its structure-guided iterative retrieval strategy for long-video understanding.

\begin{table}[t]
\centering
\small
\renewcommand{\arraystretch}{1.15}

\begin{tabular}{lcc}
\toprule
\textbf{Variants} & \textbf{VM} & \textbf{LVB} \\
\midrule
\textbf{PyraVid} & \textbf{69.1} & \textbf{58.5} \\
\midrule

\multicolumn{3}{l}{\textit{Memory Structure Ablation}} \\

\multirow{2}{*}{Plain Memory w/o Link}
& 58.7 & 53.4 \\
& {\color{red}(-15.1\%)} & {\color{red}(-8.7\%)} \\

\multirow{2}{*}{Plain Memory with Link}
& 57.1 & 54.1 \\
& {\color{red}(-17.4\%)} & {\color{red}(-7.5\%)} \\

\multirow{2}{*}{w/o global Memory w/o Link}
& 59.5 & 53.5 \\
& {\color{red}(-13.9\%)} & {\color{red}(-8.5\%)} \\

\multirow{2}{*}{w/o global Memory with Link}
& 60.3 & 54.5 \\
& {\color{red}(-12.7\%)} & {\color{red}(-6.8\%)} \\
\midrule

\multicolumn{3}{l}{\textit{Search Ablation}} \\

\multirow{2}{*}{PyraVid w/o expand (RAG)}
& 65.9 & 54.3 \\
& {\color{red}(-4.6\%)} & {\color{red}(-7.2\%)} \\

\multirow{2}{*}{PyraVid w/o prune}
& 63.5 & 57.9 \\
& {\color{red}(-8.1\%)} & {\color{red}(-1.0\%)} \\
\midrule

\multicolumn{3}{l}{\textit{Visual Ablation}} \\

\multirow{2}{*}{PyraVid w/o visual memory}
& 66.7 & 56.3 \\
& {\color{red}(-3.5\%)} & {\color{red}(-3.8\%)} \\
\bottomrule
\end{tabular}
\caption{Ablation studies on different variants of PyraVid. Relative changes compared to PyraVid are shown in parentheses.}
\label{table:components ablation}
\end{table}

\subsection{Ablation Studies}

\paragraph{Memory Structure Ablation}
Table~\ref{table:components ablation} shows that the memory structure is a core component of PyraVid. PyraVid explicitly organizes memory at different granularities and from different perspectives: fact nodes preserve grounded episodic observations, clip nodes provide higher-level semantic summaries, and global memory aggregates long-range context. When this structured design is removed, performance drops substantially. For example, replacing PyraVid with plain memory without links reduces Video-MME from 69.1 to 58.7 and LVBench from 58.5 to 53.4. Similarly, removing global memory also leads to a large performance drop, with the variant without global memory or links achieving only 59.5 on Video-MME and 53.5 on LVBench. These results show that multi-granularity memory system is important for long-horizon reasoning.

\paragraph{Search Strategy Ablation}
Table~\ref{table:components ablation} further shows that the retrieval strategy is another important component of PyraVid. Without expansion, retrieval relies only on the initially retrieved semantically similar seed nodes, making it difficult to recover supporting evidence that is not directly similar to the query. As a result, PyraVid without expansion drops to 65.9 on Video-MME and 54.3 on LVBench. Conversely, removing pruning also hurts performance, especially on Video-MME (63.5), because unfiltered expansion introduces irrelevant memory nodes into the context and increases noise. This confirms that both expansion and pruning are necessary: expansion improves evidence coverage, while pruning preserves context quality.

\paragraph{Visual Memory Ablation}
Table~\ref{table:components ablation} also shows the effect of removing visual memory from PyraVid. Removing visual memory leads to a smaller performance drop, from 69.1 to 66.7 on Video-MME and from 58.5 to 56.3 on LVBench. We believe this is related to the granularity of the evaluated questions. For many questions in these benchmarks, the key evidence is already captured by fact and clip memories, so removing explicit visual memory does not cause severe degradation. These results suggest that visual memory provides additional grounding, but its contribution is more dependent on the dataset and question type. In practice, this indicates a trade-off between accuracy and storage overhead.

\subsection{Controlled Comparison under Matched Backbone Settings}
\begin{figure}[t]
\centering
\includegraphics[width=\columnwidth]{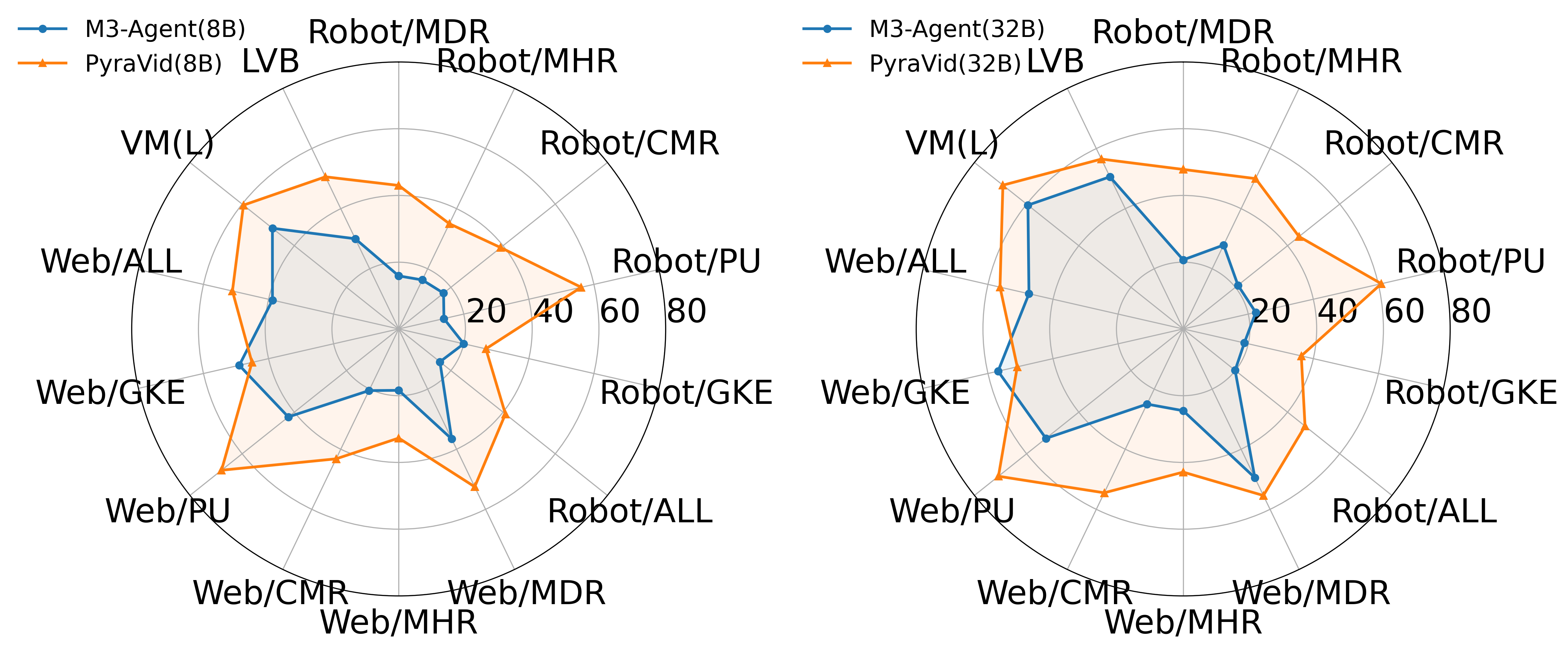}
\caption{Controlled comparison between PyraVid and M3-Agent under different matched backbone settings.}
\label{fig:matched_backbone_radar}
\end{figure}

We further conduct a controlled comparison between PyraVid and M3-Agent under matched backbone settings, as shown in Figure~\ref{fig:matched_backbone_radar}. In this setting, both frameworks use Gemini-2.0-Flash for memory construction, while answer generation is performed with either Qwen3-8B-VL-Instruct or Qwen3-32B-VL-Instruct. This design controls for the effect of backbone choice and isolates the contribution of memory organization and retrieval strategy. Across both the 8B and 32B settings, PyraVid consistently outperforms M3-Agent on nearly all benchmark dimensions, including M3-Bench-robot, M3-Bench-web, Video-MME, and LVBench. These results suggest that PyraVid’s advantage does not stem merely from stronger backbone models, but primarily from its hierarchical memory structure and its structure-guided evidence expansion and pruning mechanism. More detailed results are provided in Appendix~\ref{app:extended results}.

\subsection{Sensitivity to the Number of Seed Nodes}
\begin{figure}[t]
\centering
\includegraphics[width=\columnwidth]{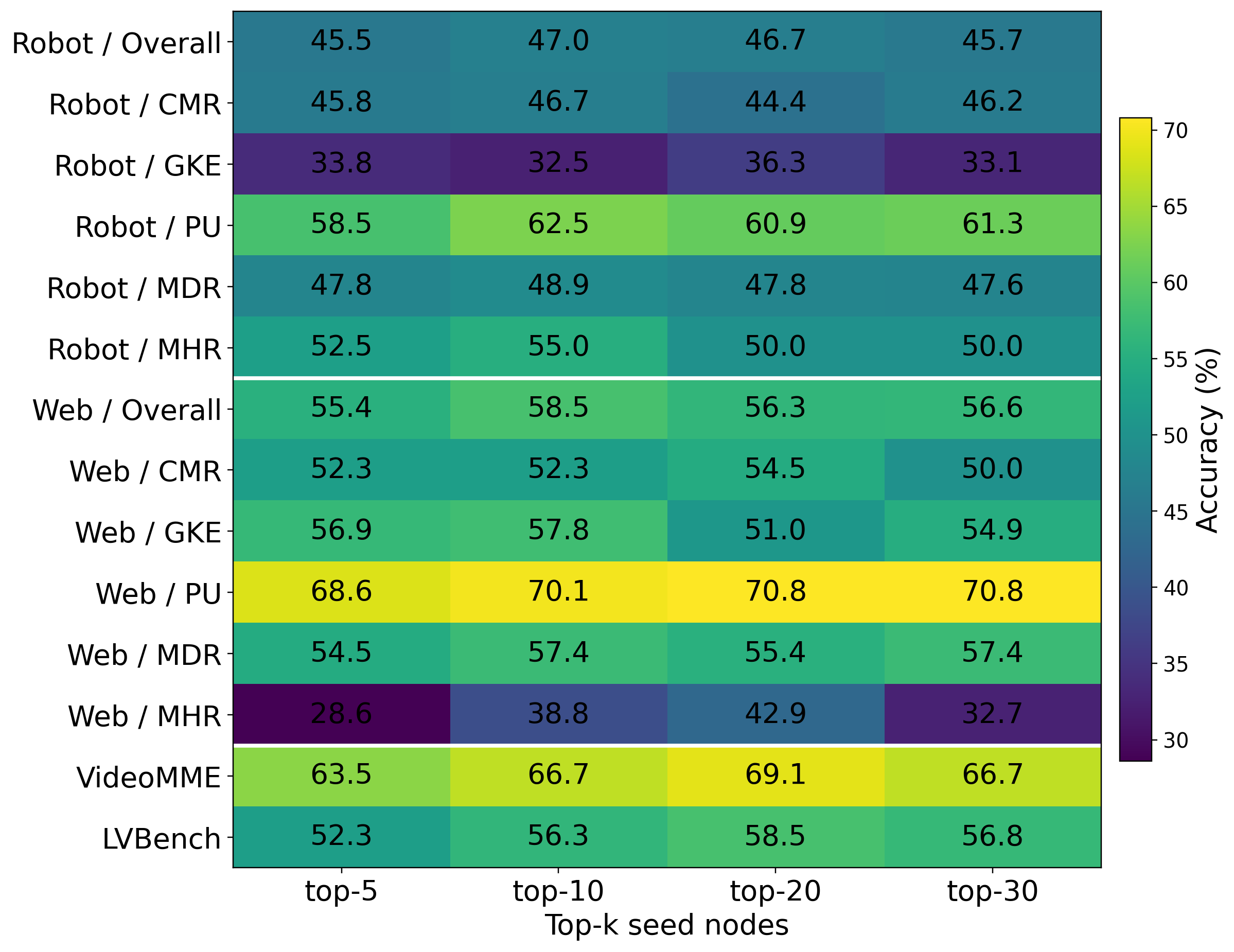}
\caption{Top-k sensitivity analysis across benchmarks and question types. Cells report performance under different numbers of seed nodes, showing that PyraVid remains broadly robust to the initial retrieval budget.}
\label{fig:topk-heatmap}
\end{figure}

Figure~\ref{fig:topk-heatmap} shows that PyraVid remains relatively stable across different choices of the initial seed size. Although the best-performing top-k varies slightly across benchmarks and question categories, moderate values such as 10 and 20 generally yield strong performance. More importantly, performance remains stable within the tested range, indicating that PyraVid is not overly sensitive to this hyperparameter. This robustness arises from structure-guided evidence expansion and pruning, which compensates for limited initial coverage when top-k is small and suppresses noisy evidence when top-k is larger.

\begin{figure}[t]
\centering
\includegraphics[width=\columnwidth]{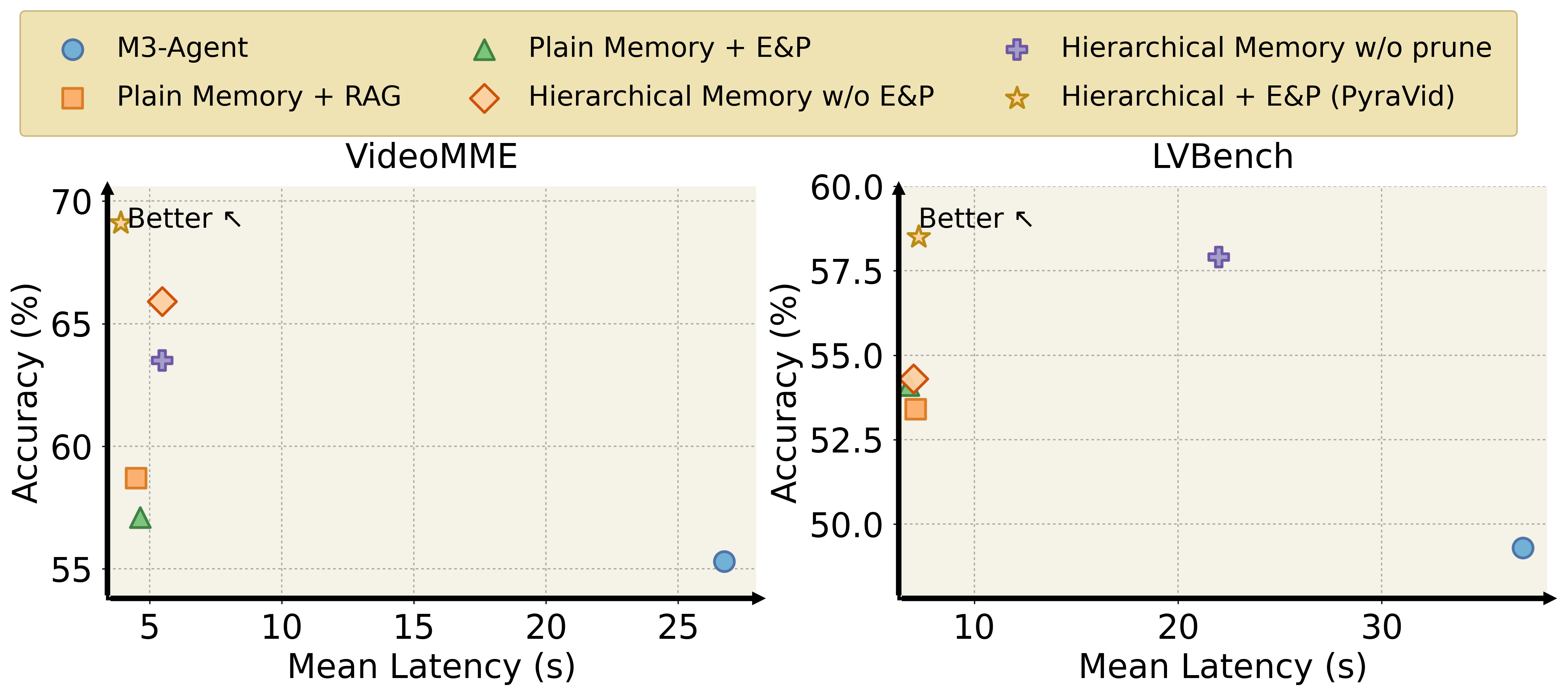}
\caption{Latency analysis on VideoMME and LVBench. Each point corresponds to a different memory design, showing the trade-off between answer accuracy and mean inference latency; points closer to the upper-left indicate better efficiency--performance balance.}
\label{fig:latency}
\end{figure}

\subsection{Latency Analysis}

Figure~\ref{fig:latency} compares answer accuracy and inference latency under different system designs. Overall, PyraVid achieves the best trade-off between effectiveness and efficiency on both benchmarks. In particular, the full \textit{Hierarchical + E\&P} design attains the highest accuracy while maintaining low latency, showing that the proposed retrieval pipeline is not only more effective but also more efficient in practice.

Among the ablated variants, removing pruning leads to the largest increase in latency. The \textit{Hierarchical Memory w/o prune} variant has a mean latency of 5.47 on VideoMME and 21.99 on LVBench. This suggests that, without pruning, expansion introduces excessively long contexts that substantially slow down answer generation. In contrast, replacing hierarchical memory with plain memory slightly reduces latency but causes a much larger drop in accuracy, suggesting that the efficiency gain does not compensate for the loss of structured evidence organization.

We also compare PyraVid with the agent-based baseline M3-Agent. PyraVid consistently outperforms M3-Agent in both accuracy and latency. We attribute the higher latency of M3-Agent to its answer-time reasoning procedure: after retrieving an initial top-k set, it further evaluates whether the current evidence is sufficient, identifies missing information, and generates additional queries for iterative agentic search. Although this design may improve flexibility, it also introduces substantial overhead. In contrast, PyraVid performs structure-guided expansion and pruning directly over the memory graph, enabling more efficient evidence aggregation with lower latency. More detailed results are provided in Appendix~\ref{app:extended results}.

\section{Conclusion}

 We presented PyraVid, a hierarchical multimodal memory framework for long-horizon video reasoning. By organizing video information into fact, clip, and global memory, and by combining structure-guided evidence expansion with pruning during inference, PyraVid enables effective aggregation of distributed evidence across different granularities and logical relations. Experiments on four benchmarks show that PyraVid consistently outperforms prior approaches, while ablation studies confirm the importance of both hierarchical memory structure and retrieval design. These results highlight the value of structured multimodal memory for scalable long-video understanding.

\clearpage
\section*{Limitations}
This work primarily studies hierarchical multimodal memory as a memory system for long-video question answering and reasoning. While this setting provides a controlled and informative testbed, it does not fully capture the broader capabilities that may be required in real-world interactive settings, such as continual learning, self-evolving, or the transfer of learned knowledge across domains and tasks. A promising direction for future work is to extend memory systems from benchmark-based reasoning toward learning in richer multimodal environments, where agents may need to acquire, consolidate, and reuse memory through ongoing perception and interaction. We believe that studying memory systems in such multimodal scenarios could further broaden the applicability of long-horizon memory systems.



\bibliography{custom}

\begin{thebibliography}{24}
\providecommand{\natexlab}[1]{#1}

\bibitem[{Bai et~al.(2025)Bai, Cai, Chen, Chen, Chen, Cheng, Deng, Ding, Gao, Ge et~al.}]{bai2025qwen3}
Shuai Bai, Yuxuan Cai, Ruizhe Chen, Keqin Chen, Xionghui Chen, Zesen Cheng, Lianghao Deng, Wei Ding, Chang Gao, Chunjiang Ge, and 1 others. 2025.
\newblock Qwen3-vl technical report.
\newblock \emph{arXiv preprint arXiv:2511.21631}.

\bibitem[{Chhikara et~al.(2025)Chhikara, Khant, Aryan, Singh, and Yadav}]{chhikara2025mem0buildingproductionreadyai}
Prateek Chhikara, Dev Khant, Saket Aryan, Taranjeet Singh, and Deshraj Yadav. 2025.
\newblock \href {https://arxiv.org/abs/2504.19413} {Mem0: Building production-ready ai agents with scalable long-term memory}.
\newblock \emph{Preprint}, arXiv:2504.19413.

\bibitem[{Comanici et~al.(2025)Comanici, Bieber, Schaekermann, Pasupat, Sachdeva, Dhillon, Blistein, Ram, Zhang, Rosen et~al.}]{comanici2025gemini}
Gheorghe Comanici, Eric Bieber, Mike Schaekermann, Ice Pasupat, Noveen Sachdeva, Inderjit Dhillon, Marcel Blistein, Ori Ram, Dan Zhang, Evan Rosen, and 1 others. 2025.
\newblock Gemini 2.5: Pushing the frontier with advanced reasoning, multimodality, long context, and next generation agentic capabilities.
\newblock \emph{arXiv preprint arXiv:2507.06261}.

\bibitem[{Fu et~al.(2025)Fu, Dai, Luo, Li, Ren, Zhang, Wang, Zhou, Shen, Zhang, Chen, Li, Lin, Zhao, Li, Xu, Zheng, Chen, Shan, He, and Sun}]{videomme}
Chaoyou Fu, Yuhan Dai, Yongdong Luo, Lei Li, Shuhuai Ren, Renrui Zhang, Zihan Wang, Chenyu Zhou, Yunhang Shen, Mengdan Zhang, Peixian Chen, Yanwei Li, Shaohui Lin, Sirui Zhao, Ke~Li, Tong Xu, Xiawu Zheng, Enhong Chen, Caifeng Shan, and 2 others. 2025.
\newblock \href {https://arxiv.org/abs/2405.21075} {Video-mme: The first-ever comprehensive evaluation benchmark of multi-modal llms in video analysis}.
\newblock \emph{Preprint}, arXiv:2405.21075.

\bibitem[{He et~al.(2024)He, Li, Jang, Jia, Cao, Shah, Shrivastava, and Lim}]{malmm}
Bo~He, Hengduo Li, Young~Kyun Jang, Menglin Jia, Xuefei Cao, Ashish Shah, Abhinav Shrivastava, and Ser-Nam Lim. 2024.
\newblock \href {https://arxiv.org/abs/2404.05726} {Ma-lmm: Memory-augmented large multimodal model for long-term video understanding}.
\newblock \emph{Preprint}, arXiv:2404.05726.

\bibitem[{Hu et~al.(2024)Hu, Chen, Chen, Mu, Shao, and Luo}]{hu2024hiagenthierarchicalworkingmemory}
Mengkang Hu, Tianxing Chen, Qiguang Chen, Yao Mu, Wenqi Shao, and Ping Luo. 2024.
\newblock \href {https://arxiv.org/abs/2408.09559} {Hiagent: Hierarchical working memory management for solving long-horizon agent tasks with large language model}.
\newblock \emph{Preprint}, arXiv:2408.09559.

\bibitem[{Li et~al.(2025{\natexlab{a}})Li, Zhang, Bo, Tian, Chen, Dai, Dong, and Tang}]{li2025camconstructivistviewagentic}
Rui Li, Zeyu Zhang, Xiaohe Bo, Zihang Tian, Xu~Chen, Quanyu Dai, Zhenhua Dong, and Ruiming Tang. 2025{\natexlab{a}}.
\newblock \href {https://arxiv.org/abs/2510.05520} {Cam: A constructivist view of agentic memory for llm-based reading comprehension}.
\newblock \emph{Preprint}, arXiv:2510.05520.

\bibitem[{Li et~al.(2024)Li, Wang, Yu, Zeng, Zhu, Huang, Gao, Li, He, Wang et~al.}]{li2024videochat}
Xinhao Li, Yi~Wang, Jiashuo Yu, Xiangyu Zeng, Yuhan Zhu, Haian Huang, Jianfei Gao, Kunchang Li, Yinan He, Chenting Wang, and 1 others. 2024.
\newblock Videochat-flash: Hierarchical compression for long-context video modeling.
\newblock \emph{arXiv preprint arXiv:2501.00574}.

\bibitem[{Li et~al.(2025{\natexlab{b}})Li, Song, Wang, Niu, Chen, Yang, Xi, Lai, Zhao, Wang, Ren, Lin, Huo, Chen, Chen, Li, Yin, Yu, Tang, Yang, Xu, and Xiong}]{li2025memosoperatingmemoryaugmentedgeneration}
Zhiyu Li, Shichao Song, Hanyu Wang, Simin Niu, Ding Chen, Jiawei Yang, Chenyang Xi, Huayi Lai, Jihao Zhao, Yezhaohui Wang, Junpeng Ren, Zehao Lin, Jiahao Huo, Tianyi Chen, Kai Chen, Kehang Li, Zhiqiang Yin, Qingchen Yu, Bo~Tang, and 3 others. 2025{\natexlab{b}}.
\newblock \href {https://arxiv.org/abs/2505.22101} {Memos: An operating system for memory-augmented generation (mag) in large language models}.
\newblock \emph{Preprint}, arXiv:2505.22101.

\bibitem[{Lin et~al.(2025)Lin, Wang, Ye, Fu, Li, and Chen}]{lin2025hippommhippocampalinspiredmultimodalmemory}
Yueqian Lin, Qinsi Wang, Hancheng Ye, Yuzhe Fu, Hai~"Helen" Li, and Yiran Chen. 2025.
\newblock \href {https://arxiv.org/abs/2504.10739} {Hippomm: Hippocampal-inspired multimodal memory for long audiovisual event understanding}.
\newblock \emph{Preprint}, arXiv:2504.10739.

\bibitem[{Long et~al.(2025)Long, He, Ye, Pan, Lin, Li, Zhao, and Li}]{m3agent}
Lin Long, Yichen He, Wentao Ye, Yiyuan Pan, Yuan Lin, Hang Li, Junbo Zhao, and Wei Li. 2025.
\newblock \href {https://arxiv.org/abs/2508.09736} {Seeing, listening, remembering, and reasoning: A multimodal agent with long-term memory}.
\newblock \emph{Preprint}, arXiv:2508.09736.

\bibitem[{Mao et~al.(2025)Mao, Perez-Cabarcas, Kallakuri, Waytowich, Lin, and Mohsenin}]{mao2025multiragmultimodalretrievalaugmentedgeneration}
Mingyang Mao, Mariela~M. Perez-Cabarcas, Utteja Kallakuri, Nicholas~R. Waytowich, Xiaomin Lin, and Tinoosh Mohsenin. 2025.
\newblock \href {https://arxiv.org/abs/2505.23990} {Multi-rag: A multimodal retrieval-augmented generation system for adaptive video understanding}.
\newblock \emph{Preprint}, arXiv:2505.23990.

\bibitem[{Rasmussen et~al.(2025)Rasmussen, Paliychuk, Beauvais, Ryan, and Chalef}]{rasmussen2025zeptemporalknowledgegraph}
Preston Rasmussen, Pavlo Paliychuk, Travis Beauvais, Jack Ryan, and Daniel Chalef. 2025.
\newblock \href {https://arxiv.org/abs/2501.13956} {Zep: A temporal knowledge graph architecture for agent memory}.
\newblock \emph{Preprint}, arXiv:2501.13956.

\bibitem[{Song et~al.(2024)Song, Chai, Wang, Zhang, Zhou, Wu, Chi, Guo, Ye, Zhang, Lu, Hwang, and Wang}]{moviechat}
Enxin Song, Wenhao Chai, Guanhong Wang, Yucheng Zhang, Haoyang Zhou, Feiyang Wu, Haozhe Chi, Xun Guo, Tian Ye, Yanting Zhang, Yan Lu, Jenq-Neng Hwang, and Gaoang Wang. 2024.
\newblock \href {https://arxiv.org/abs/2307.16449} {Moviechat: From dense token to sparse memory for long video understanding}.
\newblock \emph{Preprint}, arXiv:2307.16449.

\bibitem[{Wang et~al.(2025{\natexlab{a}})Wang, He, Hong, Cheng, Zhang, Qi, Gu, Huang, Xu, Dong, Ding, and Tang}]{lvbench}
Weihan Wang, Zehai He, Wenyi Hong, Yean Cheng, Xiaohan Zhang, Ji~Qi, Xiaotao Gu, Shiyu Huang, Bin Xu, Yuxiao Dong, Ming Ding, and Jie Tang. 2025{\natexlab{a}}.
\newblock \href {https://arxiv.org/abs/2406.08035} {Lvbench: An extreme long video understanding benchmark}.
\newblock \emph{Preprint}, arXiv:2406.08035.

\bibitem[{Wang and Chen(2025)}]{wang2025mirixmultiagentmemoryllmbased}
Yu~Wang and Xi~Chen. 2025.
\newblock \href {https://arxiv.org/abs/2507.07957} {Mirix: Multi-agent memory system for llm-based agents}.
\newblock \emph{Preprint}, arXiv:2507.07957.

\bibitem[{Wang et~al.(2025{\natexlab{b}})Wang, Takanobu, Liang, Mao, Hu, McAuley, and Wu}]{wang2025memalphalearningmemoryconstruction}
Yu~Wang, Ryuichi Takanobu, Zhiqi Liang, Yuzhen Mao, Yuanzhe Hu, Julian McAuley, and Xiaojian Wu. 2025{\natexlab{b}}.
\newblock Mem-$\{$$\backslash$alpha$\}$: Learning memory construction via reinforcement learning.
\newblock \emph{arXiv preprint arXiv:2509.25911}.

\bibitem[{Wang et~al.(2025{\natexlab{c}})Wang, Yu, Stengel-Eskin, Yoon, Cheng, Bertasius, and Bansal}]{wang2025videotree}
Ziyang Wang, Shoubin Yu, Elias Stengel-Eskin, Jaehong Yoon, Feng Cheng, Gedas Bertasius, and Mohit Bansal. 2025{\natexlab{c}}.
\newblock Videotree: Adaptive tree-based video representation for llm reasoning on long videos.
\newblock In \emph{Proceedings of the Computer Vision and Pattern Recognition Conference}, pages 3272--3283.

\bibitem[{Xu et~al.(2025)Xu, Liang, Mei, Gao, Tan, and Zhang}]{xu2025amemagenticmemoryllm}
Wujiang Xu, Zujie Liang, Kai Mei, Hang Gao, Juntao Tan, and Yongfeng Zhang. 2025.
\newblock \href {https://arxiv.org/abs/2502.12110} {A-mem: Agentic memory for llm agents}.
\newblock \emph{Preprint}, arXiv:2502.12110.

\bibitem[{Yan et~al.(2025)Yan, Yang, Huang, Nie, Ding, Li, Ma, Bi, Kersting, Pan, Schütze, Tresp, and Ma}]{yan2026memoryr1enhancinglargelanguage}
Sikuan Yan, Xiufeng Yang, Zuchao Huang, Ercong Nie, Zifeng Ding, Zonggen Li, Xiaowen Ma, Jinhe Bi, Kristian Kersting, Jeff~Z. Pan, Hinrich Schütze, Volker Tresp, and Yunpu Ma. 2025.
\newblock \href {https://arxiv.org/abs/2508.19828} {Memory-r1: Enhancing large language model agents to manage and utilize memories via reinforcement learning}.
\newblock \emph{Preprint}, arXiv:2508.19828.

\bibitem[{Yue et~al.(2026)Yue, Zhang, Peng, Fan, Guo, Li, and Zhang}]{yue2026memtdensifyingrewardslonghorizon}
Yanwei Yue, Guibin Zhang, Boci Peng, Xuanbo Fan, Jiaxin Guo, Qiankun Li, and Yan Zhang. 2026.
\newblock \href {https://arxiv.org/abs/2601.23014} {Mem-t: Densifying rewards for long-horizon memory agents}.
\newblock \emph{Preprint}, arXiv:2601.23014.

\bibitem[{Zacks et~al.(2007)Zacks, Speer, Swallow, Braver, and Reynolds}]{zacks2007event}
Jeffrey~M. Zacks, Nicole~K. Speer, Khena~M. Swallow, Todd~S. Braver, and Jeremy~R. Reynolds. 2007.
\newblock \href {https://doi.org/10.1037/0033-2909.133.2.273} {Event perception: a mind-brain perspective}.
\newblock \emph{Psychological Bulletin}, 133(2):273--293.

\bibitem[{Zhang et~al.(2024)Zhang, Wang, Tang, Liu, Feng, Dai, and Jin}]{flashvstream}
Haoji Zhang, Yiqin Wang, Yansong Tang, Yong Liu, Jiashi Feng, Jifeng Dai, and Xiaojie Jin. 2024.
\newblock \href {https://arxiv.org/abs/2406.08085} {Flash-vstream: Memory-based real-time understanding for long video streams}.
\newblock \emph{Preprint}, arXiv:2406.08085.

\bibitem[{Zhong et~al.(2023)Zhong, Guo, Gao, Ye, and Wang}]{zhong2023memorybankenhancinglargelanguage}
Wanjun Zhong, Lianghong Guo, Qiqi Gao, He~Ye, and Yanlin Wang. 2023.
\newblock \href {https://arxiv.org/abs/2305.10250} {Memorybank: Enhancing large language models with long-term memory}.
\newblock \emph{Preprint}, arXiv:2305.10250.

\end{thebibliography}

\clearpage
\appendix

\section{Case Study}
\label{app:case study}

\paragraph{Case Study: How Expansion and Pruning Lead to the Final Answer}
Figure~\ref{fig:case study expand} shows a representative reasoning trajectory of PyraVid on a long-video question: \textit{"If the woman in this video wears and changes one piece of clothes every day, then at least how many days is the video shot for?"} 

At Turn 0, PyraVid retrieves top-10 seed nodes and selects three relevant candidates. These initial memories mainly describe the overall home-renovation episode and suggest that the woman appears in a consistent outfit. Based on this limited evidence, the model cannot confidently answer the question and therefore triggers evidence expansion.

At Turn 1, PyraVid expands from the retained nodes to structurally related memories. The new evidence adds more context about the renovation process, but still does not reveal enough visual differences in clothing across scenes. The model again judges the evidence to be insufficient and continues expanding.

At Turn 2, PyraVid reaches additional fact memories from more distant segments of the video. These memories contain key visual observations showing the woman in multiple distinct outfits across scenes. By aggregating these fine-grained cues, PyraVid infers that the video spans multiple days. Combined with the broader temporal progression of the renovation project, the model finally predicts D (5 days), which is correct.

This case study demonstrates that PyraVid answers the question not by relying on a single retrieval step, but by iteratively expanding over structured links until sufficient evidence is collected. In particular, the final answer depends on combining distributed visual evidence that is not fully available in the initial seed set.

\vspace{2mm}
\begin{figure*}[htbp]
\centering
\begin{minipage}{0.9\textwidth}
\scriptsize
\begin{tcolorbox}[
promptstyle={Case Study of Iterative Evidence Expansion and Prune in PyraVid},
]
\begin{Verbatim}[breaklines=true,breakanywhere=true,  breaksymbolleft={},breaksymbolright={},breaksymbolsepleft=0pt,breaksymbolsepright=0pt]
=== Initial Node Selection ===
top-20 idx: [833, 830, 861, 992, 991, 860, 832, 831, 990, 839, 841, 842, 739, 989, 708, 497, 213, 993, 743, 494]
=== Exploration Turn 0 ===
LLM selected 3 candidates: [861, 992, 860]
LLM response: The scene immediately following Karen Sim submitting a sighting photo and earning a badge on the phone is part of a nature walk in a park or mangrove setting. eThe retrieved passage explicitly mentions that the clip includes shots of a rooster and a dragonfly during this nature walk. However, none of the options (Whale, Crocodile, Lizard, Hippo) match these animals. The passage does not mention any of the animals listed in the options. Since the context does not provide sufficient information to determine which animal is shown in the scene after the phone interaction, and the options do not align with the described wildlife (rooster, dragonfly), the evidence is incomplete.

[Expand]
[Agent] Expansion triggered.
=== Exploration Turn 1 ===
LLM selected 4 candidates: [848, 852, 860, 861]
LLM response: The scene where Karen Sim submits a sighting photo and earns a badge on the phone occurs at timestamp 42:45-42:46. The passage states that the clip shows scenes from a nature walk featuring wildlife, including a rooster and a dragonfly, and that Karen Sim and Jeffrey Chong are volunteers walking and talking on a wooden bridge. The rooster is shown at 42:00 and the dragonfly at 42:06. Since the phone submission scene is at 42:45-42:46, and the rooster and dragonfly are shown earlier in the same nature walk sequence, the next animal shown after the phone submission would logically be one of these, as they are the only animals mentioned in the context. However, the question asks for the animal shown in the scene that follows the phone submission. The provided images show the rooster and dragonfly, but no other animals are mentioned or shown in the passages or images after the phone submission. Since the rooster is shown first (42:00) and the dragonfly second (42:06), and the phone submission is at 42:45, which is after both, the next animal shown after the phone submission is not explicitly stated. But the context implies the nature walk continues, and the rooster and dragonfly are the only animals shown. Since the rooster is shown first, and the dragonfly second, and the phone submission is after both, the next animal shown after the phone submission is not specified. However, the question asks for the animal shown in the scene that follows the phone submission. Given the sequence, the rooster is shown at 42:00, dragonfly at 42:06, and phone submission at 42:45. The next scene after 42:46 is not described. But the passage says "The clip also includes shots of a rooster and a dragonfly" and the context is a nature walk. Since the rooster is shown first, and the dragonfly second, and the phone submission is after both, the next animal shown after the phone submission is not specified. However, the question asks for the animal shown in the scene that follows the phone submission. Given the options, none of them are rooster or dragonfly. Therefore, the answer must be based on the closest available information. But since the options do not include rooster or dragonfly, and the next animal is not specified, the correct action is to expand.

[Expand]
[Agent] Expansion triggered.
=== Exploration Turn 2 ===
LLM selected 1 candidates: [862]
LLM response: The scene immediately following Karen Sim submitting a sighting photo and earning a badge on the phone (at 42:46) is described in the passage at 42:47: "A monitor lizard swims in a body of water." This is a direct visual event that occurs right after the phone interaction. The monitor lizard is a type of lizard, which corresponds to option C.

[ANSWER] C
Time taken: 9.17 seconds
Question: What animal is shown in the scene that follows the scene Karen Sim submits a sighting photo and earns a badge on the phone?
Prediction: C
Gold: C
Result:  Correct
\end{Verbatim}
\end{tcolorbox}
\caption{Case study illustrating how iterative expansion over the hierarchical memory enables PyraVid to identify additional supporting evidence and arrive at the correct answer. Starting from semantically similar seed nodes, PyraVid repeatedly expands to related memories until sufficient evidence is collected.}
\label{fig:case study expand}
\end{minipage}
\end{figure*}

\section{Implementation Details}
\label{app:implementation}

Our pipeline processes long videos online to build the hierarchical memory of PyraVid. It consists of four stages: (1) memory and keyframe extraction, (2) online character memory bank construction, (3) hierarchical memory graph construction, and (4) structure-guided reasoning with expansion and pruning.

\subsection{Memory and Keyframe Extraction}

We divide each video into 30-second clips and process them independently for efficiency. Using Gemini-2.0-Flash with a predefined extraction prompt, we convert each clip into two levels of memory: fact memory and clip memory.

Fact memory captures fine-grained events within the clip and includes the following fields: \textit{description}, \textit{scene\_description}, \textit{asr}, \textit{asr\_periods}, \textit{name\_mentions}, \textit{timestamp}, and \textit{key\_frames}. Clip memory provides a coarse-grained representation of the clip, including a clip summary and an overall scene description.

We extract keyframes using the timestamps of fact memories and recover the corresponding frames with MoviePy. The extracted frames are stored as JPEG images and Base64-encoded strings for downstream use.

\subsection{Online Incremental Character Memory Bank Construction}

Based on the extracted keyframes and the fact and clip memories, we incrementally build a global character memory bank while processing clips sequentially. This stage contains five components: (1) face extraction and merging, (2) voice extraction, (3) character-level memory rewriting, (4) face--voice alignment, and (5) incremental profiling and memory update.

\paragraph{Face Extraction and Merging}
We use InsightFace (buffalo\_l) to detect faces from keyframes and encode them into embeddings. To group face identities robustly, we adopt a two-stage clustering strategy. First, we perform local clustering within each clip using HDBSCAN. Then, we incrementally merge local clusters into global identities by comparing each local centroid with existing global identity centroids using cosine similarity. If the highest similarity exceeds a threshold, the local cluster is merged; otherwise, a new global face ID is created.

\paragraph{Voice Extraction}
Using the ASR timestamps in fact memories, we extract the corresponding audio segments and encode them into embeddings for downstream matching and retrieval. Each voice segment is assigned a unique voice ID.

\paragraph{Character-Level Memory Rewriting}
To obtain character-centric memories, we align the extracted memories with detected identities. Specifically, visual face tracks and voice segments are provided to Gemini-2.0-Flash together with the original fact and clip memories. The model rewrites the textual descriptions by grounding character mentions to detected face IDs and voice IDs.
This process converts the original memories into character-level fact memories and character-level clip memories, which explicitly associate events and scene descriptions with consistent character identities.

\paragraph{Incremental Profiling and Memory Update}
After alignment, face and voice information are merged into unified person entities. For each entity, we aggregate all associated multimodal evidence, including face observations, voice segments, and related character-level fact memories.
Based on the aggregated character-level facts, Gemini-2.0-Flash incrementally builds a character profile for each person. As new clips arrive, newly extracted character-level facts are merged with the historical profile to refine the representation over time. Finally, each person entity, together with its profile and metadata, is stored in a Qdrant vector database for efficient retrieval.

\subsection{Hierarchical Memory Graph Construction}

We incrementally construct the hierarchical memory graph of PyraVid to organize multimodal video knowledge across multiple levels of abstraction.

\paragraph{Memory Node Construction}
We derive two types of memory nodes from the extracted memories: fact memory nodes and clip memory nodes, including their character-level variants. Each node stores its associated memory content as metadata, including the original descriptions, character-resolved descriptions, and clip/fact identifiers.
For retrieval, only the raw textual descriptions are used for semantic matching. Specifically, fact memory descriptions and clip memory summaries are encoded using \texttt{text-embedding-3-large}, and the resulting embeddings are stored in Qdrant for efficient similarity search. Character-level memories remain attached as auxiliary metadata and are used during reasoning.

\paragraph{Fact Memory Graph}
The fact memory graph forms the foundation of the hierarchical memory. For each new fact memory node, we retrieve the top-$K$ most semantically similar historical fact memory nodes from Qdrant using fact memory embeddings. These candidates are then passed to Qwen3-4B-Instruct, which determines whether structured links should be established based on temporal, semantic, and logical relations.

\paragraph{Clip Memory Graph}
Each clip is represented as a clip memory node. Two types of links are constructed. First, hierarchical links connect each clip memory node to its associated fact memory nodes. Second, cross-clip links connect clip memory nodes whose underlying fact memory nodes are linked in the fact memory graph. In this way, lower-level factual relations induce higher-level structural links.

\paragraph{Global Memory Node}
At the highest level, we maintain a global memory node that stores an evolving summary of the video. As each clip is processed, Gemini-2.0-Flash updates the global memory by integrating the previous global summary with the newly generated clip memory.

\subsection{Structure-Guided Reasoning with Expansion and Pruning}

During inference, PyraVid performs structure-guided reasoning over the hierarchical memory with iterative expansion and pruning. Given a query, the system first retrieves the top-$k$ most semantically similar fact memory nodes as the initial seed nodes. The retrieved nodes are then passed to Qwen3-8B or Qwen3-32B for pruning. Starting from the retained nodes, the system identifies mentioned characters from the character-level fact and clip memories, and retrieves the corresponding character profiles from the character memory bank as supplementary context. In addition, the retained nodes, their associated keyframes, and the global memory are included as input context. These inputs are jointly provided to Qwen3-VL-8B-Instruct or Qwen3-VL-32B-Instruct for multimodal reasoning and answer generation. The model determines whether the current evidence is sufficient to answer the query. If so, it outputs the final answer. Otherwise, it performs an expansion step to retrieve additional evidence. During expansion, the retained nodes are treated as seed nodes, and the system explores related memory nodes in the hierarchical memory graph, including directly linked neighbors. If a seed node is a clip memory node, all fact memory nodes belonging to that clip are also added to the candidate pool. The expanded nodes are merged with the retained nodes, and the updated evidence context enters the next cycle of pruning, reasoning, and expansion. This process continues until the model determines that the evidence is sufficient or a maximum number of expansion steps is reached.

\subsection{Baseline Implementation}
\paragraph{Socratic Model Baselines}
We segment each video into clips and build a verbal-based memory bank by prompting a vision-language model to summarize every clip with a short paragraph describing the main actions, objects, people, scene, and visible or spoken text. These clip descriptions are embedded and stored in a per-video Qdrant vector database. At inference time, each question is embedded and used to retrieve the top-20 most relevant clip memories, which are concatenated as context for answer inference. For multiple-choice questions, an LLM is prompted to predict exactly one option label; for open-ended questions, it generates a short free-form answer based on the retrieved memory. 

\begin{algorithm}[t]
\caption{Socratic Memory for Video Question Answering}
\label{alg:socratic_memory_compact}
\small
\KwIn{Video clips $\mathcal{C}$, per-video questions $\mathcal{Q}$, multimodal model $f$, embedding model $f_{\mathrm{emb}}$, retrieval size $k$}
\KwOut{Predicted answers}

\ForEach{video $v$}{
    initialize memory store $\mathcal{M}_v$\;
    
    \tcp{Memory construction}
    \ForEach{clip $c_i \in \mathcal{C}_v$}{
        $d_i \leftarrow f(c_i)$\;
        $\mathrm{Store}(\mathcal{M}_v, f_{\mathrm{emb}}(d_i), d_i)$\;
    }
    
    \tcp{Question answering}
    \ForEach{question $q \in \mathcal{Q}_v$}{
        $\Gamma \leftarrow \mathrm{Retrieve}(\mathcal{M}_v, f_{\mathrm{emb}}(q), k)$\;
        
        \uIf{$q$ is multiple-choice}{
            $\hat{a} \leftarrow f(q, \Gamma, \mathrm{options}(q))$\;
        }
        \Else{
            $\hat{a} \leftarrow f(q, \Gamma)$\;
        }
        
        save prediction $\hat{a}$\;
    }
}
\end{algorithm}

\section{Prompts}
\label{app:prompts}

\begin{figure*}[htbp]
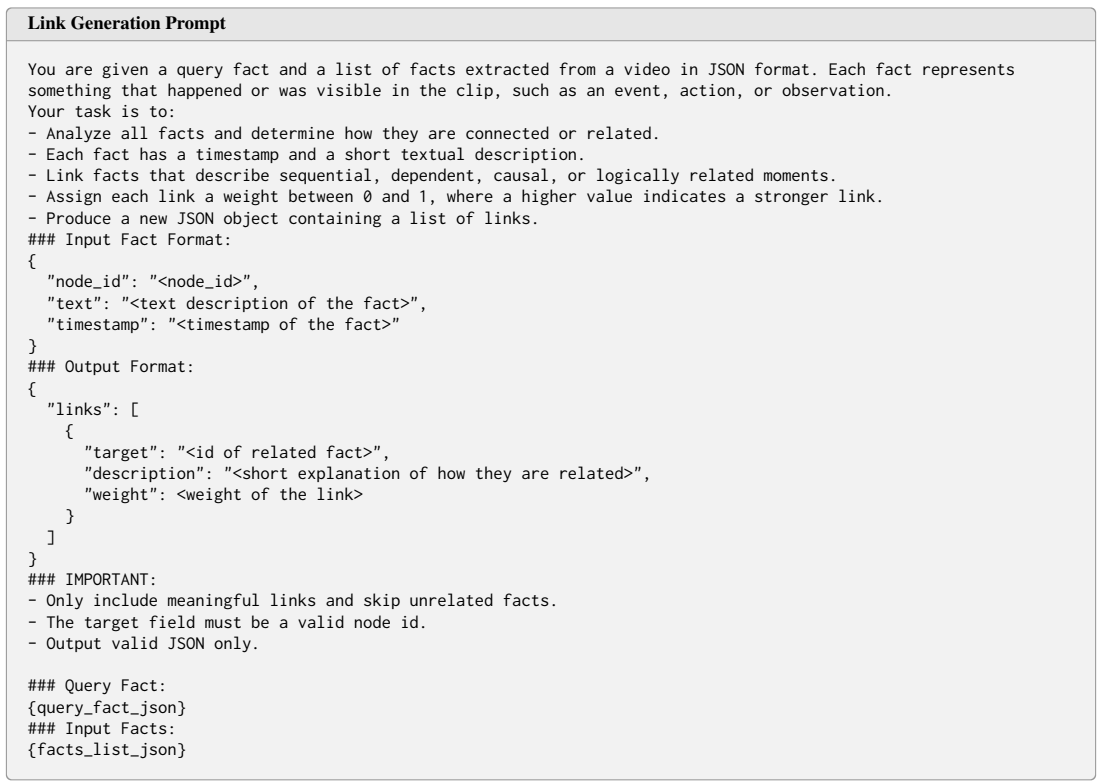

\centering
\begin{minipage}{0.9\textwidth}
\scriptsize
\begin{tcolorbox}[
  promptstyle={Link Generation Prompt},
]
\begin{verbatim}
You are given a query fact and a list of facts extracted from a video in JSON format. Each fact represents 
something that happened or was visible in the clip, such as an event, action, or observation.
Your task is to:
- Analyze all facts and determine how they are connected or related.
- Each fact has a timestamp and a short textual description.
- Link facts that describe sequential, dependent, causal, or logically related moments.
- Assign each link a weight between 0 and 1, where a higher value indicates a stronger link.
- Produce a new JSON object containing a list of links.
### Input Fact Format:
{
  "node_id": "<node_id>",
  "text": "<text description of the fact>",
  "timestamp": "<timestamp of the fact>"
}
### Output Format:
{
  "links": [
    {
      "target": "<id of related fact>",
      "description": "<short explanation of how they are related>",
      "weight": <weight of the link>
    }
  ]
}
### IMPORTANT:
- Only include meaningful links and skip unrelated facts.
- The target field must be a valid node id.
- Output valid JSON only.

### Query Fact:
{query_fact_json}
### Input Facts:
{facts_list_json}
\end{verbatim}
\end{tcolorbox}

\caption{Prompt used for relational link generation among fact memories.}
\label{fig:link_generation_prompt}
\end{minipage}
\end{figure*}

\begin{figure*}[htbp]
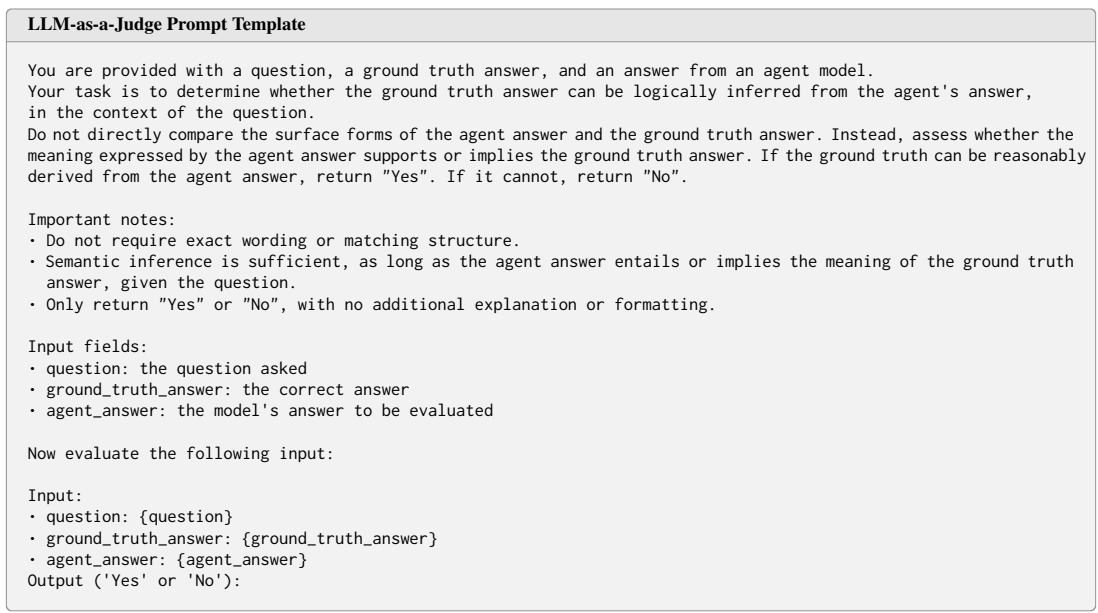

\centering
\begin{minipage}{0.9\textwidth}
\scriptsize
\begin{tcolorbox}[
  promptstyle={LLM-as-a-Judge Prompt Template},
]\begin{verbatim}
You are provided with a question, a ground truth answer, and an answer from an agent model. 
Your task is to determine whether the ground truth answer can be logically inferred from the agent's answer,
in the context of the question.  
Do not directly compare the surface forms of the agent answer and the ground truth answer. Instead, assess whether the 
meaning expressed by the agent answer supports or implies the ground truth answer. If the ground truth can be reasonably 
derived from the agent answer, return "Yes". If it cannot, return "No".  

Important notes: 
• Do not require exact wording or matching structure. 
• Semantic inference is sufficient, as long as the agent answer entails or implies the meaning of the ground truth 
  answer, given the question. 
• Only return "Yes" or "No", with no additional explanation or formatting.  

Input fields: 
• question: the question asked 
• ground_truth_answer: the correct answer 
• agent_answer: the model's answer to be evaluated  

Now evaluate the following input:
  
Input: 
• question: {question} 
• ground_truth_answer: {ground_truth_answer} 
• agent_answer: {agent_answer}  
Output ('Yes' or 'No'):
\end{verbatim}
\end{tcolorbox}
\caption{Prompt used for LLM-as-a-Judge evaluation with GPT-4o-mini.}
\label{fig:llm_as_judge_prompt}
\end{minipage}
\end{figure*}

PyraVid uses several prompts to support memory construction and evaluation. 
Figure~\ref{fig:link_generation_prompt} shows the prompt used to generate relational links among fact memories, which instructs the language model to identify sequential, causal, or logical relations between events. 
Figure~\ref{fig:llm_as_judge_prompt} presents the prompt template used for LLM-as-a-Judge evaluation, where GPT-4o-mini determines whether the predicted answer semantically entails the ground-truth answer.
Figure~\ref{fig:mm_prompt_part1} presents the prompt template used in the agentic exploration pipeline for multiple-choice questions, where the agent either answers the question directly or executes an Expand action to explore additional nodes in the memory graph.
Figure~\ref{fig:mm_prompt_part2} presents the prompt template used in the agentic exploration pipeline for open-ended questions, which follows the same reasoning mechanism. In addition, it also incorporates the relevant character information retrieved from the character database, enabling the agent to leverage identity-aware context when generating answers.
Figure~\ref{fig:node_selection_part1} presents the prompt template used for the node selection stage in multiple-choice reasoning, where the agent filters out unrelated nodes from the retrieved candidates before further exploration.
Figure~\ref{fig:node_selection_part2} presents the prompt template used for node selection in open-ended questions, which follows the same reasoning mechanism. The prompt also incorporates the relevant character information retrieved from the character database to support identity-aware reasoning.

\vspace{2mm}

\vspace{2mm}

\vspace{2mm}
\begin{figure*}[htbp]
\centering
\begin{minipage}{0.9\textwidth}
\scriptsize

\begin{tcolorbox}[promptstyle={Multiple-Choice Answering Prompt Template}]
\begin{verbatim}
You will be given:
- A multiple-choice question about a video
- Four candidate options (A, B, C, D)
- A general context summary
- Retrieved passages
- Available images

You must decide between Two possible actions:

-------------------------------------------------------
Option 1 — Answer

If the current context is sufficient and you are confident:

- Reason over the context, passages, question, and options carefully.
- Please first give a brief reasoning and immediately after the reasoning, output a space and then:
  [ANSWER] X
- X must be exactly one uppercase letter: A, B, C, or D.
- Stop immediately after "[ANSWER] X".
- Do NOT write anything else.
- When conflicting evidence exists, pick the most direct visual evidence and STOP reasoning.

Example:
The woman is holding the trophy in the final scene. [ANSWER] A

-------------------------------------------------------
Option 2 — Expand

If the current evidence is relevant but incomplete,
and expanding nearby graph nodes may help:

Output exactly:
[Expand]

Nothing else.

Important Rules:

- Choose exactly ONE action.
- Do NOT output explanations outside the required format.
- If you have uncertainty, choose [Expand] rather than guessing.
- If format is violated, the answer will be discarded.

Inputs:
Question: {question}
Context Summary: {context_summary}
Options: {options}
Passages: {passages}

Output:
\end{verbatim}
\end{tcolorbox}

\caption{Prompt used for multiple-choice questions.}
\label{fig:mm_prompt_part1}
\end{minipage}
\end{figure*}

\vspace{2mm}
\begin{figure*}[htbp]
\centering
\begin{minipage}{0.9\textwidth}
\scriptsize
\begin{tcolorbox}[promptstyle={Open Question Answering Prompt Template}]
\begin{verbatim}
You will be given:
- An open question about a video
- A general context summary
- Retrieved passages
- Character profiles referenced in the passages using tags such as <person_1>.
- Available images

You must decide between Two possible actions:

-------------------------------------------------------
Option 1 — Answer

If the current context is sufficient and you are confident:

- Please output the final answer in the exact format specified below.:
  [ANSWER] ....
  ....should be the answer text, not just a letter, and it should be as concise as possible.
- Stop immediately after "[ANSWER] ....".
- Do NOT write anything else.
- When conflicting evidence exists, pick the most direct visual evidence and STOP reasoning.
- Do not contain any person identifiers (e.g., <person_1>) in the final answer.

Example:
From the text given and images, the woman is holding the trophy in the final scene. 
[ANSWER] The woman is holding the trophy.

-------------------------------------------------------
Option 2 — Expand

If the current evidence is relevant but incomplete,
and expanding nearby graph nodes may help:

Output exactly:
[Expand]

Nothing else.

Important Rules:

- Choose exactly ONE action.
- Do NOT output explanations outside the required format.
- If you have uncertainty, choose [Expand] rather than guessing.
- If format is violated, the answer will be discarded.

Inputs:
Question: {question}
Context Summary: {context_summary}
Passages: {passages}
Character Profiles: {character_profiles}

Output:
\end{verbatim}
\end{tcolorbox}

\caption{Prompt used for open questions.}
\label{fig:mm_prompt_part2}
\end{minipage}
\end{figure*}

\vspace{2mm}
\begin{figure*}[htbp]
\centering
\begin{minipage}{0.9\textwidth}
\scriptsize
\begin{tcolorbox}[
  promptstyle={Multiple-Choice Node Selection Prompt Template},
]
\begin{verbatim}
You will be provided with:

1. A question about a video.
2. A general high-level summary of the video.
3. A set of extracted passages, each being either:
   - an atomic fact (with a single timestamp), or
   - a clip-level summary (with a time range) which indicates the contents of its underlying facts.

Your task is to select **all passages that contain information potentially helpful** for answering the question.  
A passage is helpful if it directly answers, partially answers, or provides relevant context for the question.

Passages are formatted as a JSON as the following:
{{
  // fact passage
  <passage_number>: {{
    "text": "<fact_text>",
    "timestamp": "<timestamp>",
  }}
  // clip-level summary passage
  <passage_number>: {{
    "text": "<summary_text>",
    "timestamp_start": "<timestamp_start>",
    "timestamp_end": "<timestamp_end>",
  }}
  ...
}}

Return **only** a list of the passage numbers you deem helpful.  
Example output: [1, 3, 5]
Please do not include any extra text or explanation.

**Question:**
{question}

**General Summary of the Video:**
{context_summary}

**Video Passages:**
{passages}

**A LIST containing all helpful passage numbers:**
\end{verbatim}
\end{tcolorbox}
\caption{Prompt used for multiple-choice node selection.}
\label{fig:node_selection_part1}
\end{minipage}
\end{figure*}

\vspace{2mm}
\begin{figure*}[htbp]
\centering
\begin{minipage}{0.9\textwidth}
\scriptsize
\begin{tcolorbox}[promptstyle={Open Question Node Selection Prompt Template}]
\begin{verbatim}
You will be provided with:

1. A question about a video.
2. A general high-level summary of the video.
3. A set of extracted passages, each being either:
   - an atomic fact (with a single timestamp), or
   - a clip-level summary (with a time range) which indicates the contents of its underlying facts.
4. Character profiles referenced in the passages using tags such as <person_1>.

Your task is to select **all passages that contain information potentially helpful** for answering the question.  
A passage is helpful if it directly answers, partially answers, or provides relevant context for the question.

Passages are formatted as a JSON as the following:
{{
  // fact passage
  <passage_number>: {{
    "text": "<fact_text>",
    "timestamp": "<timestamp>"
  }}
  // clip-level summary passage
  <passage_number>: {{
    "text": "<summary_text>",
    "timestamp_start": "<timestamp_start>",
    "timestamp_end": "<timestamp_end>"
  }}
  ...
}}

Character profiles are formatted as a JSON as following:
{{
  "<person_id>": "<character_profile_text>",
  ...
}}

Return **only** a list of the passage numbers you deem helpful.  
Example output: [1, 3, 5]
Please do not include any extra text or explanation.

**Question:**
{question}

**General Summary of the Video:**
{context_summary}

**Character Profiles:**
{character_profiles}

**Video Passages:**
{passages}

**A LIST containing all helpful passage numbers:**
\end{verbatim}
\end{tcolorbox}
\caption{Prompt used for open question node selection.}
\label{fig:node_selection_part2}
\end{minipage}
\end{figure*}

\section{Extended Results}
\label{app:extended results}
For completeness, we provide the numerical results corresponding to Figure~\ref{fig:matched_backbone_radar} and Figure~\ref{fig:latency} from the main paper in Table~\ref{table:controlled comparison} and Table~\ref{tab:latency_accuracy_mc_mr}, respectively. These tables present the same comparisons in tabular form for easier reading and more precise value inspection.
Table~\ref{table:controlled comparison} reports the controlled comparison between PyraVid and M3-Agent under matched backbone settings. Table~\ref{tab:latency_accuracy_mc_mr} reports answer accuracy and inference latency on VideoMME and LVBench, where latency is summarized by the p50, p95, and mean values.

\begin{table*}[t]
\centering
\footnotesize
\setlength{\tabcolsep}{2pt}
\renewcommand{\arraystretch}{1.2}

\resizebox{\textwidth}{!}{
\begin{tabular}{lcccccc cccccc cc}
\toprule
\multirow{2}{*}{\textbf{Model}} &
\multicolumn{6}{c}{\textbf{M3-Bench-robot}} &
\multicolumn{6}{c}{\textbf{M3-Bench-web}} &
\multirow{2}{*}{\textbf{VM(L)}} &
\multirow{2}{*}{\textbf{LVB}}  \\
\cmidrule(lr){2-7}\cmidrule(lr){8-13}
& \textbf{MDR} & \textbf{MHR} & \textbf{CMR} & \textbf{PU} & \textbf{GKE} & \textbf{ALL}
& \textbf{MDR} & \textbf{MHR} & \textbf{CMR} & \textbf{PU} & \textbf{GKE} & \textbf{ALL}
& & \\
\midrule
M3-Agent(8B)       & 15.9 & 16.3 & 17.2 & 13.9 & 20.0 & 15.8 & 36.6 & 18.4 & 20.5 & 42.3 & 49.0 & 38.8 & 48.4 & 30.0 \\
M3-Agent(32B)       & 20.7 & 27.9 & 21.0 & 22.3 & 18.8 & 19.9 & 49.5 & 24.5 & 25.0 & 52.6 & 56.9 & 47.4 & 57.1 & 41.0 \\
M3-Agent(32B RL)       & 32.8 & 29.4 & 31.2 & 43.3 & 19.1 & 30.7 & 45.9 & 28.4 & 44.3 & 59.3 & 53.9 & 48.9 & 55.3 & 49.3  \\
PyraVid(8B)        & 43.0 & 35.0 & 39.1 & 56.1 & 26.8 & 40.9 & 52.5 & 32.7 & 43.2 & 67.9 & 45.1 & 51.1 & 59.5 & 50.6 \\
PyraVid(32B)        & 47.8 & 50.0 & 44.4 & 60.9 & 36.3 & 46.7 & 55.4 & 42.9 & 54.5 & 70.8 & 51.0 & 56.3 & 69.1 & 58.5\\
\bottomrule
\end{tabular}
}
\caption{Controlled Comparison with M3-Agent under Matched Backbones}
\label{table:controlled comparison}
\end{table*}

\begin{table*}[t]
\centering
\small
\setlength{\tabcolsep}{4pt}
\begin{tabular}{l|cccc|cccc}
\hline
\multirow{2}{*}{Method} 
& \multicolumn{4}{c|}{VideoMME} 
& \multicolumn{4}{c}{lvbench} \\
& Acc (\%) $\uparrow$ & p50 $\downarrow$ & p95 $\downarrow$ & mean $\downarrow$
& Acc (\%) $\uparrow$ & p50 $\downarrow$ & p95 $\downarrow$ & mean $\downarrow$\\
\hline
M3-Agent 
& 55.3 & 20.43 & 52.61 & 26.75 & 49.3 & 36.97 & 59.24 & 36.92\\
Plain Memory + RAG
& 58.7 & 3.15 & 14.00 & 4.48 & 53.4 & 5.48 & 12.73 & 7.12\\
Plain Memory + E\&P 
& 57.1 & 2.76 & 16.34 & 4.64 & 54.1 & 3.09 & 20.70 & 6.79\\
Hierarchical Memory w/o E\&P
& 65.9 & 3.87 & 8.84 & 5.47 & 54.3 & 5.53 & \textbf{12.44} & \textbf{7.02}\\
Hierarchical Memory w/o prune 
& 63.5 & 3.04 & 16.92 & 5.47 & 57.9 & 3.70 & 105.99 & 21.99\\
\textbf{Hierarchical + E\&P (PyraVid)} 
& \textbf{69.1} & \textbf{2.22} & \textbf{7.52} & \textbf{3.90}
& \textbf{58.5} & \textbf{2.97} & 18.85 & 7.26 \\
\hline
\end{tabular}

\caption{Comparison of answer performance and inference latency on VideoMME and LVBench. Higher accuracy indicates better answer quality, while lower latency indicates higher efficiency. Latency is reported in seconds.}
\label{tab:latency_accuracy_mc_mr}
\end{table*}

\end{document}